\documentclass[letter]{jpsj2} 

\title{Three-dimensional electronic structure of superconducting iron
pnictides observed by angle-resolved photoemission spectroscopy}

\author{Walid \textsc{MALAEB}$^{1}$\thanks{malaeb@wyvern.phys.s.u-tokyo.ac.jp},
Teppei \textsc{YOSHIDA}$^{1,2}$, Atsushi \textsc{FUJIMORI}$^{1}$,
Masato \textsc{KUBOTA}$^{3}$, Kanta \textsc{ONO}$^{3}$, Kunihiro
\textsc{KIHOU}$^{4}$, Parasharam M. \textsc{SHIRAGE}$^{4}$,\\
Hijiri \textsc{KITO}$^{4}$, Akira \textsc{IYO}$^{2,4}$, Hiroshi
\textsc{EISAKI}$^{2,4}$, Yasuyuki \textsc{NAKAJIMA}$^{2,5}$,
Tsuyoshi \textsc{TAMEGAI}$^{2,5}$, and Ryotaro
\textsc{ARITA}$^{2,5}$.}

\inst{
$^{1}$Department of Complexity Science and Engineering and Department of Physics, University of Tokyo, Bunkyo-ku, Tokyo 113-0033\\
$^{2}$JST, Transformative Research-Project on Iron Pnictides (TRIP), Chiyoda, Tokyo 102-0075 \\
$^{3}$Institute for Material Structure Science, High Energy Accelerator Research Organization, Tsukuba, Ibaraki 305-0801\\
$^{4}$National Institute of Advanced Industrial Science and Technology (AIST), Tsukuba, Ibaraki 305-8568\\
$^{5}$Department of Applied Physics, University of Tokyo,
Bunkyo-ku, Tokyo 113-8656}

\abst{We have performed an angle-resolved photoemission
spectroscopy (ARPES) study of the undoped and electron-doped iron
pnictides BaFe$_{2-x}$Co$_x$As$_2$ (Ba122) ($\textit{x}$=0, 0.14)
and studied the Fermi surfaces (FSs) and band dispersions near the
Fermi level. The FS sheets we observed are consistent with the
shrinkage of the hole-like pockets around the Brillouin Zone (BZ)
center and the expansion of the electron pockets around the BZ
corner in the electron-doped compound as compared to the undoped
parent compound. Band dispersions and FSs around the BZ center
strongly depend on the photon energy, indicating the
three-dimensional (3D) electronic structure. This observation
suggests that the antiferromagnetism and superconductivity in the
pnictides may have to be considered including the
orbital-dependent 3D electronic structure, where FS nesting is not
necessarily strong.}

\kword{ pnictides, three-dimensional electronic structure, ARPES}

\begin{document}
\maketitle The discovery of superconductivity (SC) in layered iron
pnictides \cite{ref1} with the critical temperature $T_c$ reaching
$\sim$56 K \cite{ref2} has opened a new route for the high-$T_c$
research in addition to that of the cuprates, bringing new
challenges to the materials science community on both experimental
and theoretical sides. This new class of iron-based systems share
some common properties with the cuprates such as the layered
crystal structures \cite{ref1} and antiferromagnetic (AFM)
ordering in the parent compounds \cite{ref3,ref4}. However, many
differences exist between the two families especially in their
electronic structures. These differences started to appear from
the early stage when local-density-approximation (LDA)
band-structure calculations predicted that all the Fe $3d$-derived
bands exist near the Fermi level ($E_{\rm F}$), resulting in
complicated hole- and electron-like Fermi surface (FS) sheets
\cite{ref5,ref6,ref7}, whereas only a single band with one FS
(hole- or electron- like) exists in the cuprates. The predictions
of the LDA calculations were confirmed by photoemission
experiments, which demonstrated that Fe $3d$ states are
predominant near $E_{\rm F}$ \cite{ref8,ref9,ref10} with moderate
$p-d$ hybridization and electron correlation \cite{ref9}. More
detailed features of the electronic structure came from
angle-resolved photoemission spectroscopy (ARPES): (i) Several
disconnected hole-like and electron-like FS sheets \cite{ref11},
(ii) kinks in the dispersions, suggesting coupling of
quasiparticles to Boson excitations \cite{ref12,ref13}, (iii)
moderately renormalized energy bands due to electron correlation
\cite{ref14,ref15}, and (iv) FS-dependent nodeless,
nearly-isotropic superconducting gaps
\cite{ref16,ref17,ref18,ref19}. Most of these results give
reasonable agreement with the band-structure calculations.
Nevertheless, many other features of the electronic structure,
predicted by the calculations and necessary for understanding the
occurrence of SC in the iron pnictides, have not been observed yet
by ARPES.

Among these features is the three-dimensional (3D) electronic
structure of $\textit{A}$Fe$_{2}$As$_2$ ($\textit{A}$=Alkaline
earth) (122 family) and its doped compounds predicted by
band-structure calculations \cite{ref20,ref21,ref22} as compared
to the quasi-two-dimensional electronic structure of the cuprates.
Such a three-dimensionality has been supported by recent
experimental reports such as the weak anisotropy of the upper
critical field $H_{\rm c2}$ in (Ba,K)Fe$_{2}$As$_2$ \cite{ref23}
and Ba(Fe,Co)$_{2}$As$_2$ \cite{ref24',ref24} (Ba122). The weak
anisotropy of $H_{\rm c2}$ was attributed to the 3D FSs in these
compounds. Here, we show a direct observation of the 3D electronic
structure in iron pnictides by ARPES. Our results on the AFM
parent compound BaFe$_{2}$As$_2$ and the superconducting compound
BaFe$_{1.86}$Co$_{0.14}$As$_2$ show that the FS in the Brillouin
Zone (BZ) center ($\Gamma$ point) has strong modulation along the
$k_{\rm z}$ direction, whereas the FS at the BZ corner (X point)
is almost cylinder-like with weak $k_{\rm z}$ modulation
especially in the undoped sample. This is in good agreement with
the band-structure calculations \cite{ref20,ref21,ref22} and the
$H_{\rm c2}$ measurements \cite{ref23,ref24',ref24}. The
three-dimensionality of the FS around the $\Gamma$ point was
reported by two recent ARPES studies \cite{ref25,ref26},
consistent with the present results. However, in Ref. [27] the
three-dimensionality was reported to exist only in the AFM state
below the structural/magnetic transition temperature ($T_{\rm S}$)
of the undoped compound. Our results show that the superconducting
Ba(Fe,Co)$_{2}$As$_2$ sample has similarly strong three
dimensionality, yielding important implications for the mechanism
of SC as well as for the origin of the antiferromagnetism.

High-quality single crystals of the parent compound
BaFe$_{2}$As$_2$ and the electron-doped compound
BaFe$_{1.86}$Co$_{0.14}$As$_2$ ($T_c$=24 K) were grown using the
flux method \cite{ref27}. ARPES measurements were carried out at
BL-28A of Photon Factory (PF) using a circularly-polarized light
with photon energies ranging between 35 and 90 eV. A Scienta
SES-2002 analyzer was used with the total energy resolution of
$\sim$ 20 meV and the momentum resolution of $\sim$ 0.02$\pi/a$,
where $\textit{a}$=3.9\AA\ is the in-plane lattice constant. The
crystals were cleaved \textit{in situ} at $\textit{T}$=10 K in an
ultra-high vacuum below 1$\times10^{-10}$ Torr giving flat
mirror-like surfaces which stayed reasonably stable all over our
measuring time ($\sim$ 3 days). Calibration of the $E_{\rm F}$ of
the samples was achieved by referring to that of gold. We have
performed a density functional calculation with the
local-density-approximation (LDA) by using the WIEN2k package
\cite{ref27'}, where the experimental tetragonal lattice
parameters of room temperature BaFe$_{2}$As$_2$ were used. As for
the internal coordinate of As, Ref. [20] was followed and $z_{As}$
was set to be 0.342, which was obtained by LDA total energy
minimization.

\begin{figure}[tb]
\begin{center}
\includegraphics[width=180mm]{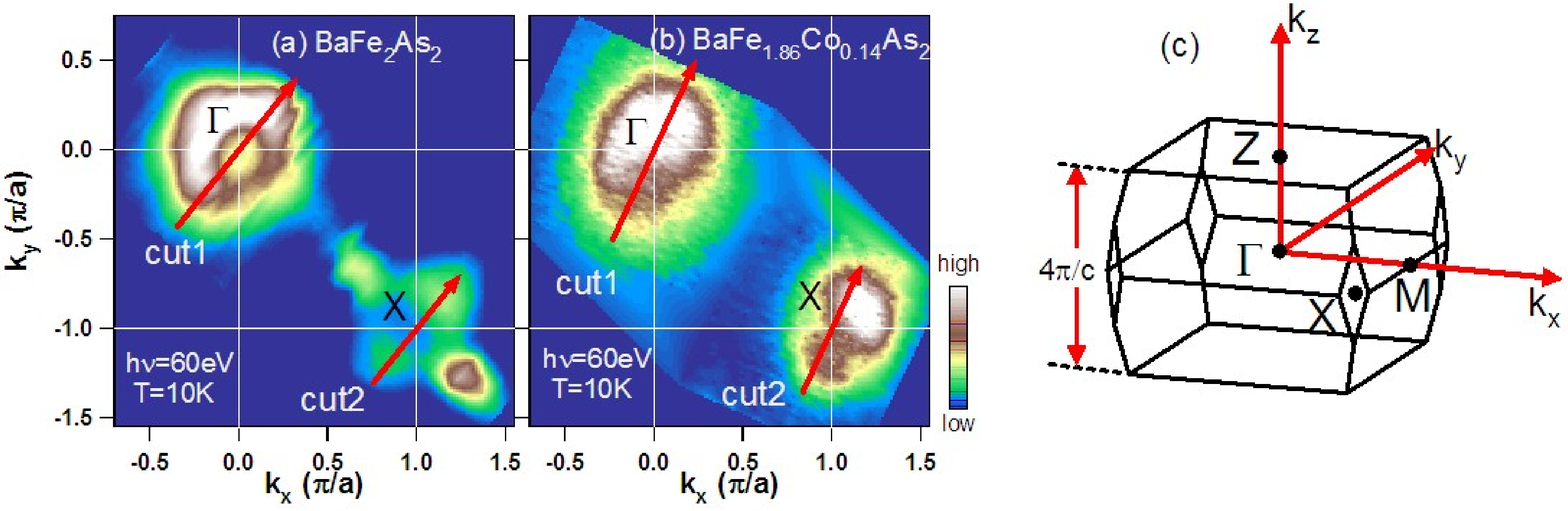}
\end{center}
\caption{(Color online) Fermi surface mapping at T=10 K and
h$\nu$=60 eV in the $k_{\rm x}$-$k_{\rm y}$ plane. The intensity
has been integrated within a window of 20 meV around $E_{\rm F}$
(a): BaFe$_{2}$As$_2$. (b): BaFe$_{1.86}$Co$_{0.14}$As$_2$. Red
arrows indicate the directions of the cuts for which the results
are shown in Figs.  \ref{f2} and {3}. (c): Three-dimensional
Brillouin zone of BaFe$_{2-x}$Co$_x$As$_2$ in the tetragonal
phase.} \label{f1}
\end{figure}

Figure \ref{f1} (a) and (b) show the results of FS mapping at low
temperature ($\sim$10 K) using photon energy $h\nu=$ 60 eV for the
undoped and electron-doped Ba122 compounds, respectively. In these
plots, the photoemission intensity has been integrated over 20 meV
around $E_{\rm F}$. For both compounds, one can clearly observe a
nearly circular-shaped hole pocket centered at the two-dimensional
(2D) BZ center (denoted by $\Gamma$) in addition to an electron
pocket centered at the 2D BZ corner (denoted by X), in agreement
with the previous ARPES results on $\textit{A}$Fe$_{2}$As$_2$
($\textit{A}$=Alkaline earth) and their doped compounds
\cite{ref16,ref18}. The 3D BZ of BaFe$_{2}$As$_2$ in the
tetragonal phase is shown in Fig. \ref{f1} (c) and is used as a
reference to indicate high-symmetry points in momentum
($\textit{k}$) space. Note that the intensity asymmetry observed
around $\Gamma$ and X is due to the photoemission matrix-element
effect arising form the circularly-polarized light used in this
study. Direct comparison between the two compounds reveals that
the hole (electron) pocket becomes smaller (larger) in size with
electron doping. Although it shrinks in size, the hole-like FS
around $\Gamma$ does not completely disappear in this case as was
observed for more heavily electron-doped Ba(Fe,Co)$_{2}$As$_2$
\cite{ref28}. The features around the X point, which are
anisotropic in the undoped compound, are expanded, change their
shapes, and become nearly circular with electron doping.

\begin{figure}[tb]
\begin{center}
\includegraphics[width=170mm]{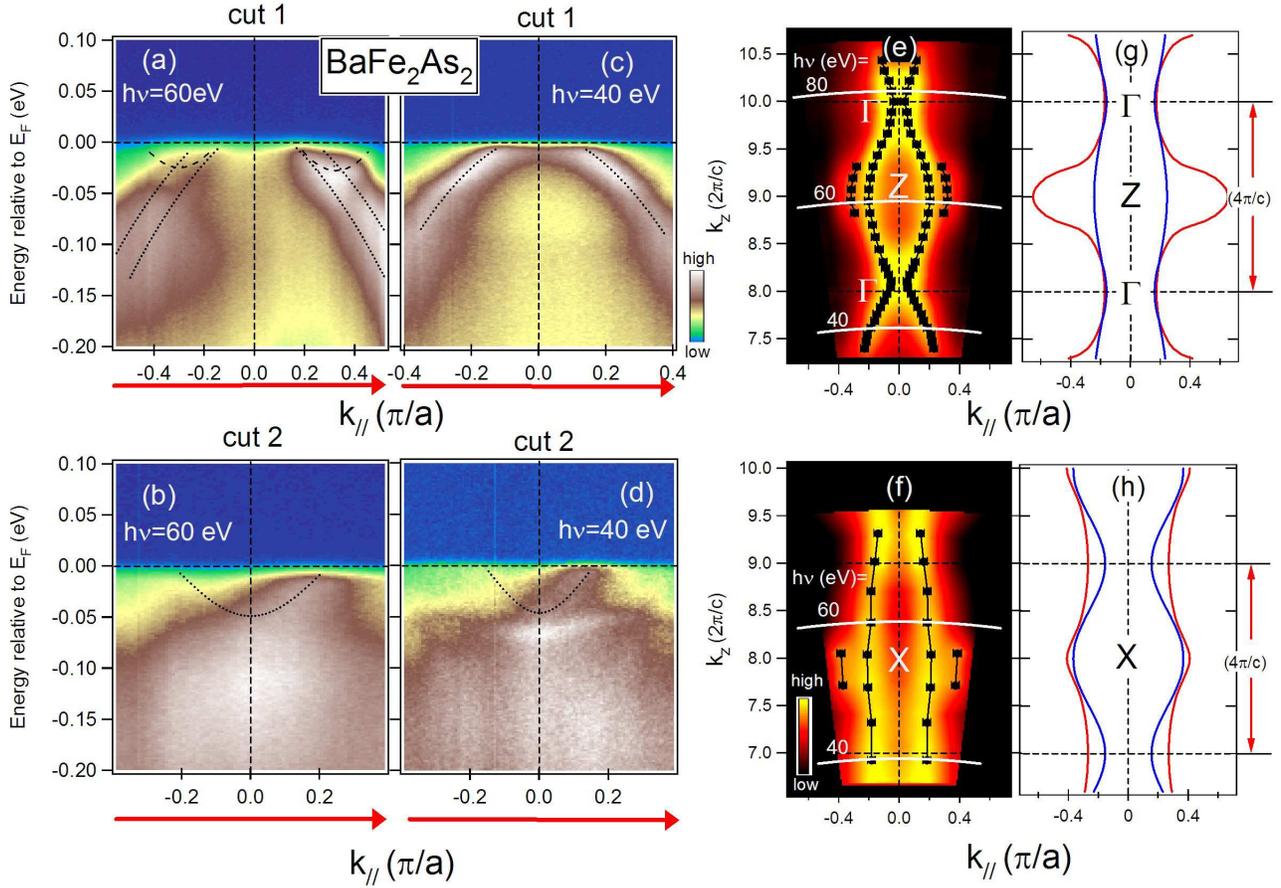}
\end{center}
\caption{(Color online) ARPES $\textit{E-k}$ intensity maps for
BaFe$_{2}$As$_2$ taken at $\textit{T}$=10 K. (a), (c): Along cut 1
in Fig.\ref{f1}(a). (b), (d): Along cut 2 in Fig. \ref{f1} (a).
The photon energies are indicated inside each panel. The black
curves are guides to the eye. The dashed curves indicate an
electron pocket-like feature. (e), (f): Fermi surface images of
BaFe$_{2}$As$_2$ in the $k_{\rm
\parallel}-k_{\rm z}$ plane obtained from the $h\nu$-dependent ARPES data
taken, respectively, along cut 1 and cut 2 in Fig. \ref{f1} (a).
The photoemission intensities have been integrated in a window of
20 meV around $E_{\rm F}$ and finally symmetrized about
$\textit{k}$=0. Both inner and outer black dots represent the
$k_{\rm F}$ values determined from MDC peak positions at $E_{\rm
F}$. (g), (h): Fermi surfaces of BaFe$_{2}$As$_2$ around the
$\Gamma$ and X points, respectively, determined by LDA
band-structure calculations with $k_{\rm \parallel}$ parallel to
the X-Z line and hence approximately to cut 1 and cut 2 in Fig.
\ref{f1} (a).}\label{f2}
\end{figure}

The ARPES intensity plot in energy-momentum ($\textit{E-k}$) space
along cuts 1 (across $\Gamma$) and 2 (across X) for undoped Ba122
taken at $h\nu=$ 60 eV are shown, respectively, in panels (a) and
(b) of Fig. \ref{f2}. In the figure, $k_{\rm \parallel}$ denotes
momenta along cut 1 or cut 2.  Panel (a) shows a hole-like band
crossing $E_{\rm F}$ and giving rise to the hole-like FS around
$\Gamma$ [Fig. \ref{f1} (a)]. However, a tiny structure having the
features of an electron pocket is observed in this case. This has
been interpreted as the result of band folding due to the AFM
order also observed in a previous ARPES study of another parent
compound of the 122 family, SrFe$_{2}$As$_2$ \cite{ref29}. We note
that recent results of quantum oscillation measurements of the
parent compound BaFe$_{2}$As$_2$ support a conventional band
folding picture of the AFM ground state of this compound
\cite{ref30}. As for cut 2 in Fig. \ref{f1} (a) (X point), an
electron-like band crossing $E_{\rm F}$ is observed.

\begin{figure}[tb]
\begin{center}
\includegraphics[width=170mm]{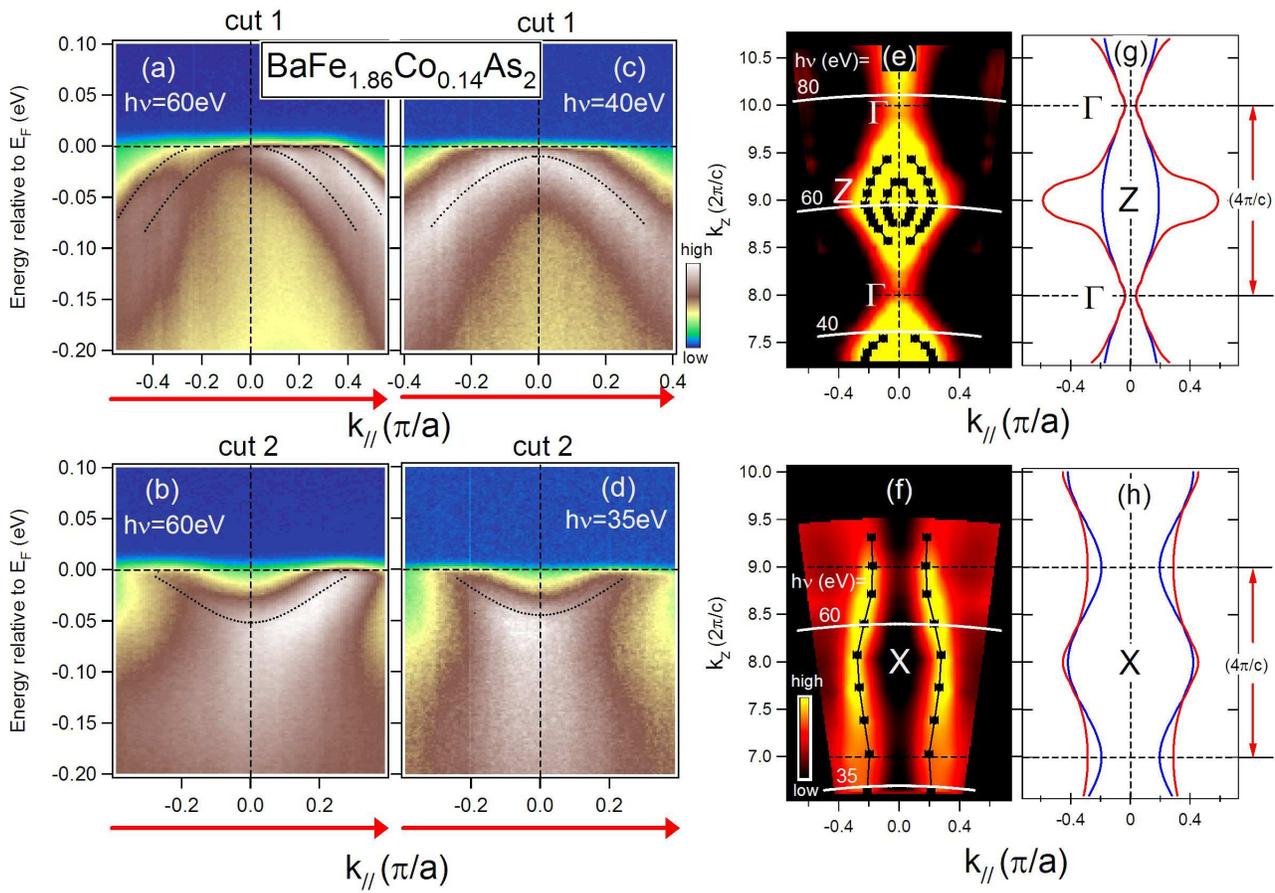}
\end{center}
\caption{(Color online) The same as Fig. \ref{f2} but for
BaFe$_{1.86}$Co$_{0.14}$As$_2$ along cut 1 and cut 2 in Fig.
\ref{f1} (b).}\label{f3}
\end{figure}

Similar cuts (cuts 1 and 2 in Fig. \ref{f1} (b)) were taken for
the electron-doped sample BaFe$_{1.86}$Co$_{0.14}$As$_2$ under the
same conditions as those of the undoped sample and the
corresponding $\textit{E-k}$ plots for $h\nu=$ 60 eV  are shown in
Fig. \ref{f3} (a) and (b). Panel (a) shows two hole-like bands
(inner and outer) crossing $E_{\rm F}$ without signature of the
additional electron pocket feature observed for the same cut in
the undoped sample [Fig. \ref{f2} (a)]. This suggests that
electron doping weakens the band folding effects arising from the
AFM order. The band leading to the electron-like FS is observed
more clearly around X [Fig. \ref{f3} (b)]. It should be noted that
the dispersions of BaFe$_{1.86}$Co$_{0.14}$As$_2$ in Fig. \ref{f3}
show some broadening as compared to those of the parent compound
BaFe$_{2}$As$_2$ (Fig. \ref{f2}), probably because Co doping
increases carrier scattering in the FeAs plane.

Now, we turn to the investigation of possible three-dimensionality
in the electronic structure ($k_{\rm z}$ dependence) predicted by
most of the band-structure calculations on iron pnictides, in
particular on 122 compounds \cite{ref20,ref21,ref22}. These
calculations predict different $k_{\rm z}$ dependences at
different regions in the $k_{\rm x}$-$k_{\rm y}$ space, namely, a
stronger three-dimensionality is expected around the 2D BZ center
$\Gamma$ (or the $\Gamma$-Z line in the 3D BZ) rather than the 2D
zone corner X (Refer to Fig. \ref{f1} (c)). It is well known that
by changing the photon energy in photoemission experiment, one can
probe different $k_{\rm z}$ values \cite{ref31} (within the
uncertainty of $k_{\rm z}$ $\approx$ 1/$\lambda$, where $\lambda$
is the photoelectron mean-free path) according to the following
formula which relates the wave vector along the
$\textit{z}$-direction ($k_{\rm z}$) to the excitation energy
\textit{h}$\nu$ for normal emission:
\begin{eqnarray}
k_{z}=\left[\frac{2m}{\hbar^2}\right]^{1/2}[(h\nu-\Phi)\cos^2\theta+V_{0}]^{1/2},
\end{eqnarray}
where $\textit{m}$ and $2\pi\hbar$ are the electron mass and
Planck's constant, \textit{h}$\nu$ is the photon energy, $\Phi$ is
the work function, and \textit{V}$_{0}$ is the inner potential of
the sample.

Thus, we performed a photon-energy dependent ARPES measurement of
both samples at two specific regions of $\textit{k}$-space, around
the $\Gamma$ point or around the $\Gamma$-Z line (cut 1 in Fig.
\ref{f1} (a) and (b)) and around the X point (cut 2 in Fig.
\ref{f1} (a) and (b)). For both samples, the ARPES intensity plots
in $\textit{E-k}$ space show strong photon-energy dependence along
the $\Gamma$-Z line. This can be clearly observed by comparing the
dispersions in Fig. \ref{f2} (a) and (c) for the undoped sample,
where the Fermi level crossing occurs at different $k_{\rm F}$
values depending on the photon energy (\textit{h}$\nu$=60 and 40
eV are shown here). This suggests that the corresponding FS
centered at the $\Gamma$ point shows strong photon-energy
dependence and consequently strong $k_{\rm z}$ dependence as
mentioned above. From a large set of photon-energy dependent ARPES
data (with photon energies ranging from 35 to 90 eV) and using Eq.
(1), we have constructed the FS images in the
$k_{\rm\parallel}$-$k_{\rm z}$ plane containing the $\Gamma$-Z
line for the undoped sample in Fig. \ref{f2} (e). The color scale
in this figure represents the photoelectron intensity integrated
in a narrow energy window of 20 meV around $E_{\rm F}$. $k_{\rm
F}$ points determined from peak positions in momentum distribution
curves (MDCs) are also shown in this plot as black dots. From this
plot it becomes clear that the FS shows strong modulation
attaining its smallest size around the $\Gamma$ point
[\textbf{k}=(0,0,0)] and its largest size around the Z point
[\textbf{k}=(0,0,2$\pi$/c)]. As for the photon-energy dependence
around the X point, Fig. \ref{f2} (d) shows the dispersion around
the X point for the undoped Ba122 sample (cut 2 in Fig. \ref{f1}
(a)) at \textit{h}$\nu$=40 eV, which when compared with the same
data taken at \textit{h}$\nu$=60 eV [Fig. \ref{f2} (b)] shows only
small differences. The FS image in the $k_{\rm
\parallel}$-$k_{\rm z}$ plane is constructed as shown in Fig.
\ref{f2} (f). The image plot shows almost straight cylinders along
the $k_{\rm z}$ direction and thus weak $k_{\rm z}$ dependence,
which renders the electronic structure around the BZ corner much
less three-dimensional than that observed around the BZ center.

Now we turn to the case of the electron-doped superconducting
sample: Fig. \ref{f3} (a) and (c) which correspond to cut 1 in
Fig. \ref{f1} (b), show a clear photon energy dependence of the
dispersions as well as $k_{\rm F}$ points. The corresponding FS
image plot along the $\Gamma$-Z line is shown in Fig. \ref{f3}
(e), where the strong $k_{\rm z}$ modulation is quite clear. The
dispersions around the X point show weaker photon-energy
dependence as can be deduced from those taken at
\textit{h}$\nu$=60 and 35 eV and shown in Fig. \ref{f3} (b) and
(d), respectively. However, the modulation along $k_{\rm z}$ can
be clearly seen in the FS image plot in Fig. \ref{f3} (f). The
three-dimensionality of this FS is stronger than that observed for
the undoped sample at the X point [Fig. \ref{f2} (f)].

These experimental data are compared with the FSs determined from
our LDA band-structure calculations in the paramagnetic state in
panels (g) and (h) of Figs. \ref{f2} and \ref{f3}. Because the
major effect of AFM ordering is to fold the energy bands into the
AFM BZ and resultant band anticrossing accompanied by intensity
redistribution, the global intensity distribution is not
dramatically altered by the band folding, which is also confirmed
by calculations due to the small magnetic moments in pnictides
\cite{ref32}. We thus find qualitatively good agreement between
experiment and calculation although the calculation predicts
generally stronger three-dimensionality. We note that the FSs
derived from the \textit{$d_{z^2}$} orbitals have stronger $k_{\rm
z}$ modulation than others. Another discrepancy between experiment
and calculation is that the experiment does not necessarily reveal
the second FSs (except for some photon energies). In order to
observe the second FSs, further experiment with higher statistics
and using various photon polarizations will be necessary.

Observing the 3D electronic structure in the parent compound
BaFe$_{2}$As$_2$, which shows a collinear AFM spin density wave
(SDW) \cite{ref4}, raises the question about the role of FS
nesting in driving the AFM state in the parent compounds
\cite{ref6}. Our experimental data and LDA calculations show that
the outer hole-like FS sheet around $\Gamma$ has stronger
three-dimensionality as compared to the electron-like FS sheet
around X. It seems that the strong three-dimensionality
significantly weakens the nesting condition between the two FS
sheets, which was proposed to enhance the high-$T_c$ SC in this
compound \cite{ref16,ref18,ref28}. Now we discuss the relationship
between the three-dimensional electronic structure and SC in iron
pnictides. In this class of materials, the 122 family is predicted
to have more pronounced three-dimensionality than the 1111 family
\cite{ref33} by the LDA band-structure calculations
\cite{ref20,ref21,ref22}. This may be correlated with the
different $T_c$ values, the highest $T_c$ ($\sim$56K) among the
pnictides until now being achieved in the 1111 family \cite{ref2}
as compared to $\sim$38K in the 122 family \cite{ref34}. It has
also been observed by ARPES that the size of the SC gap is larger
in the 1111 family (for example $\Delta\sim$15 meV in
NdFeAsO$_{0.9}$F$_{0.1}$) \cite{ref17} than in the 122 family (for
example $\Delta\sim$12 meV in Ba$_{0.6}$K$_{0.4}$Fe$_2$As$_2$
\cite{ref16} and $\Delta\sim$6.7 meV in
BaFe$_{1.85}$Co$_{0.15}$As$_2$ \cite{ref18}). Nevertheless, the
fairly high-$T_c$ values attained by the 122 family suggest that
the high-$T_c$ SC exists not only in systems with low-dimensional
electronic structures like cuprates. This is also consistent with
the results from $H_{\rm c2}$ measurements in the 122 system
mentioned above \cite{ref23, ref24', ref24}. According to the LDA
band-structure calculations [Fig.  \ref{f3} (g)], the outer
hole-like FS sheet is derived mainly from \textit{$d_{z^2}$}
orbitals. In the scenario of spin-fluctuation-mediated SC, the 2D
\textit{$x^{2}-y^{2}/xy$} bands play major roles in SC and the
\textit{$d_{z^2}$} band is only a "passive" band for SC. The small
gap $\Delta$ for these bands may be regarded as the "passiveness"
of the \textit{$d_{z^2}$} band but may not be so "passive" since
2$\Delta$/k$_{B}T_{c}$ is comparable to the active 2D FS in the
1111 compounds. Judging the real contribution of the 3D electronic
structure to the SC in iron pnictides needs further investigation.

A recent ARPES report on CaFe$_{2}$As$_2$ \cite{ref26} has shown
that its FSs have strong three-dimensionality, consistent with the
present study but that a 3D to 2D transition occurs above the
orthorhombic-to-tetragonal structural/magnetic transition
temperature ($T_S$= 160 $\sim$ 170 K). While this led the authors
to the conclusion that the low dimensionality plays an important
role in understanding the superconducting mechanism in pnictides,
our results on the superconducting sample
BaFe$_{1.86}$Co$_{0.14}$As$_2$ confirm that high-$T_c$ SC exists
in pnictides in the three-dimensional electronic structure.
Raising the temperature above $T_S$ and Co-doping in
\textit{A}Fe$_{2}$As$_2$ destroy the AFM ordering and lead to the
orthorhombic-to-tetragonal structural transition. However, this
does not necessarily imply that a similar 3D to 2D transition
occurs with carrier doping. As for the case of Co-doping, our
ARPES results on superconducting BaFe$_{1.86}$Co$_{0.14}$As$_2$
confirm that the three-dimensionality persists in the SC samples,
and that achieving relatively high-$T_c$ SC in iron pnictides with
3D electronic structure is possible although the relationship
between three-dimensionality and SC needs further clarification in
future studies.

In summary, we have performed an ARPES study on the undoped and
electron-doped iron pnictides Ba(Fe,Co)$_2$As$_2$ and studied the
FSs and band dispersions near $E_{\rm F}$. The strong
photon-energy dependence of band dispersions and FSs around the BZ
center of both the parent and superconducting compounds indicates
the 3D electronic structure, where FS nesting should be weakened
compared with the 2D electronic structure. These observations
suggest that the antiferromagnetism and the AFM ordering and
high-$T_c$ SC in the pnictides may have to be considered including
the 3D electronic structure.\\ \\\ \ \ \

The authors acknowledge A. Ino, Y. Aiura, Y. Nakashima, Y. Ishida,
T. Shimojima and S. Shin for informative discussions and S. Ideta
for technical support. This work was supported by a Grant-in-Aid
for Scientific Research in Priority Area "Invention of Anomalous
Quantum Materials" from the Ministry of Education, Culture,
Sports, Science and Technology (MEXT) and by Japan Science and
Technology Agency (JST). WM is thankful to MEXT for financial
support. Experiment at Photon Factory was approved by the Photon
Factory Program Advisory Committee (Proposal No. 2006S2-001).

\newpage

\end{document}